\documentclass[twocolumn,showpacs,preprintnumbers,amsmath,amssymb]{revtex4}
\usepackage{color}
\usepackage{bm, graphicx, amsmath}
\usepackage{hyperref}

\begin{document}

\title{Retrodiction of a sequence of measurement results in qubit interferometers}
\author{Mark Hillery and Daniel Koch}
\affiliation{Department of Physics, Hunter College of the City University of New York, 695 Park Avenue, New York, NY 10065 USA \\ and Physics Program, Graduate Center of the City University of New York, 365 Fifth Avenue, New York, NY 10016}

\begin{abstract}
We study how well we can retrodict results of measurements made on a quantum system if we can make measurements on its final state.  We know what measurements were made, but not their results.  An initial examination shows that we can gain anywhere from no information to perfect information about the results of previous measurements, depending on the measurements and the initial state of the system.  The case of two two-outcome measurements, the second of which is a projective measurement, is examined in some detail.  We then look at a model of a qubit interferometer in which measurements are made in order to determine the path the qubit followed.  The measurement made on the final state of the qubit depends on the information about previous measurement results that we are trying to determine.  One can attempt to find the result of just one of the measurements, all of them, or find a measurement sequence that was not realized.  We study all three possibilities.
\end{abstract}  


\maketitle

\section{Introduction}
How much can you tell about the past of a quantum system from its present state?  This is the problem of retrodiction \cite{pegg1,barnett1,barnett2,pegg2,barnett3}.  One is often concerned with prediction, for example, describing the results of measurements made on the system at a later time.  Retrodiction is concerned with  the past of the system.   An example of a situation that encompasses both prediction and retrodiction is given by the standard setup in quantum communication theory \cite{barnett3}.  Alice chooses a state from a set of states known to Alice and Bob, and sends it to Bob, who then measures the state.  Alice would like to predict the result of Bob's measurement based on which state she sent, and Bob would like to retrodict which state Alice sent.

The theory of quantum retrodiction allows one to define a retrodictive state, which can be used to make predictions \cite{pegg1}.  This point of view has been used to analyze a number of systems in quantum optics, including a beam splitter \cite{pegg1}, amplifiers and attenuators \cite{barnett2}, and a driven atom \cite{pegg2}.  It can be applied to both closed and open systems \cite{pegg2}.

Here we shall be interested in the retrodiction of measurement results.  Suppose a quantum system has been prepared in a quantum state and then subjected to a series of measurements.  Different measurement results will lead to different final states of the system.  We assume that all we have access to is the final state of the system, and not the results of the measurements, and we would like to gain information about those results.  The set of measurement results can be viewed as a trajectory of the quantum system, and we will explore what can be learned about that trajectory from the final state of the system.  We may be interested in only part of the trajectory, the entire trajectory, or determining whether a particular trajectory did not occur.

The problem studied here is closely related to that of sequential measurements on the same quantum system \cite{rapcan,bergou,heinosaari}.  Rather surprisingly, it has been shown that one can gain information about the initial state of a system even though a measurement has intervened and changed the state of the system \cite{rapcan,bergou}.  In our case, for the measurements determining the trajectory, subsequent measurements can disturb the quantum state resulting from a previous one, thereby complicating the task of determining the trajectory.

We shall approach the problem of retrodicting measurement results from the final state in two ways.  After a short discussion of some simple cases, we will see what can be done when a quantum system is subjected to two two-outcome measurements, the second of which is a projective measurement. Next, we will study a simple model that will allow us to look at more general types of measurements.  The picture behind the model is that of a photon going through a sequence of interferometers, where in each interferometer there is a detector that gives us information about which path the photon took through that interferometer.  We would like to find out what we can infer about the photon's path, i.e.\ the results of the path detectors, from its state when it emerges from the final interferometer.  We will make use of a qubit instead of a photon, and instead of measuring paths, our detectors will tell us whether the qubit is in the state $|0\rangle$ or $|1\rangle$. 

\section{Some simple cases}
A measurement is described by a Positive Operator Valued Measure (POVM), which is a set of positive operators, $\{ \Pi_{j}=A^{\dagger}_{j}A_{j} | j=1,2,...n\}$ such that $\sum_{j=1}^{n}\Pi_{j}=I$.  If the state being measured is $\rho$, then the probability of obtaining the result $j$ is $p_{j}={\rm Tr}(\Pi_{j}\rho )$, and if the result $j$ is obtained, the state after the measurement is $A_{j}\rho A^{\dagger}_{j}/p_{j}$.  While this not the most general measurement model possible, (see \cite{hein}), it will suffice for our purposes here. 

We can obtain an idea of the range of possible relations between a sequence of measurement results and the final state of a system by considering some simple examples.  At one extreme, there are cases in which we learn nothing about the measurement results from the final state of the system.  Let us consider making two measurements on a qubit,  with the first measurement described by $\{ \Pi_{aj}=A^{\dagger}_{j}A_{j} | j=0,1\}$ the second by $\{ \Pi_{bj}=B^{\dagger}_{j}B_{j} | j=0,1\}$.  We will consider the case in which all of these operators are diagonal in the computational basis $\{ |0\rangle ,|1\rangle \}$,
\begin{equation}
A_{j}=\left(\begin{array}{cc} a_{j0} & 0 \\ 0 & a_{j1} \end{array}\right) \hspace{5mm}
B_{j}=\left(\begin{array}{cc} b_{j0} & 0 \\ 0 & b_{j1} \end{array}\right) ,
\end{equation}
where 
\begin{eqnarray}
\sum_{j=0}^{1} |a_{j0}|^{2}=\sum_{j=0}^{1} |a_{j1}|^{2}=1 \\
\sum_{j=0}^{1} |b_{j0}|^{2}=\sum_{j=0}^{1} |b_{j1}|^{2}=1 .
\end{eqnarray}
Now suppose we start the qubit in the state $|0\rangle$.  The probability that we obtain $j$ for the first measurement and $k$ for the second, where $j,k\in \{ 0,1\}$ is $|a_{j0}|^{2} |b_{k0}|^{2}$, but in all cases the final state of the system is $|0\rangle$.  Therefore, in this case, we learn nothing about the results of the measurements from the final state of the system.  We also note that each measurement is independent of the ones before it.

A less extreme case is when the measurement operators are one-dimensional projections. Then the final state of the system is determined only by the final measurement result, and so it would seem to carry no information about the previous ones.  However, the probability that a particular final state occurs does depend on the results of the previous measurements, so we can infer some information about those measurements from the final state.  A measurement sequence of this type can be described as a Markov chain.  The probability of a measurement result only depends on the result of the previous measurement, because that measurement determines the state that is being measured.

Finally, suppose our system consists of two qubits, and the measurement operators are given by $A_{j} = P_{j}\otimes I$ and $B_{j}=I\otimes P_{j}$, where $P_{0}$ and $P_{1}$ are orthogonal one-dimensional projections.  The first measurement only measures the first qubit, and the second measures the second.  In this case, different final states are correlated with different sequences of measurement results, and these states are orthogonal.  Therefore, by measuring the final state of the system we will know what both measurement results were.

What we can conclude from these examples is that there is wide range of behaviors possible.  Correlations between final states and measurement results can range from nonexistent to perfect.  In order to further examine what is possible, let us first look at the case of two two-outcome measurements.

\section{Two two-outcome measurements}
We start with the system in the state $|\psi\rangle$, and perform two two-outcome measurements on it.  We denote the outcomes of the measurements by $\{ +,-\}$.  The first measurement is described by a POVM $\Pi_{\pm} =A_{\pm}^{\dagger}A_{\pm}$, where $\Pi_{+}+\Pi_{-}=I$.  If the measurement result was $+$ the post-measurement state is $A_{+}|\psi\rangle /\|A_{+}\psi \|$, and if the result was $-$ it is $A_{-}|\psi\rangle /\|A_{-}\psi \|$.  We shall assume for now that the second measurement is described by the projections $Q_{\pm}$, where $Q_{+}+Q_{-}=I$.

Now suppose we have been given the system after the measurements have been made, and we would like to determine the result of the first measurement.  This can be viewed as a problem of discriminating between two density matrices.  The first density matrix is the one that results at the output if the result of the first measurement was $+$, which is given by
\begin{equation}
\rho_{1+}=\frac{1}{p_{1+}}[Q_{+}A_{+}|\psi\rangle\langle\psi |A_{+}^{\dagger}Q_{+}+ Q_{-}A_{+}|\psi\rangle\langle\psi |A_{+}^{\dagger}Q_{-}] ,
\end{equation}
and it occurs with a probability of $p_{1+}=\langle\psi |\Pi_{+}|\psi\rangle$.  The second density matrix is the one that results if the result of the first measurement is $-$,
\begin{equation}
\rho_{1-}=\frac{1}{p_{1-}}[Q_{+}A_{-}|\psi\rangle\langle\psi |A_{-}^{\dagger}Q_{+}+ Q_{-}A_{-}|\psi\rangle\langle\psi |A_{-}^{\dagger}Q_{-}] ,
\end{equation}
and it occurs with a probability of $p_{1-}=\langle\psi |\Pi_{-}|\psi\rangle$.  

These density matrices cannot, in general, be perfectly distinguished, so we need to turn to a strategy that will give us some information about which one we have.  The minimum-error strategy minimizes the probability of making a mistake.  Suppose we are trying to discriminate between two density matrices, $\rho_{a}$, which occurs with probability $p_{a}$, and $\rho_{b}$, which occurs with probability $p_{b}$.  Minimum-error discrimination gives us a two-element POVM, $\{\Pi_{a},\Pi_{b}\}$, where $\Pi_{a}$ corresponds to detecting $\rho_{a}$ and $\Pi_{b}$ corresponds to detecting $\rho_{b}$.  The probability of successfully identifying the state is
\begin{equation}
P_{s}=p_{a} {\rm Tr}(\rho_{a}\Pi_{a}) + p_{b}{\rm Tr}(\rho_{b}\Pi_{b}) ,
\end{equation}
and for the optimal POVM, that is, for the one that minimizes the probability of making a mistake, is given by
\begin{equation}
\label{min-err}
P_{s}=\frac{1}{2}[1 + \|p_{a}\rho_{a}-p_{b}\rho_{b}\|_{1} ] ,
\end{equation}
where the norm in the above equation is the trace norm \cite{helstrom}.  Setting $\Lambda = p_{a}\rho_{a}-p_{b}\rho_{b}$, the POVM element corresponding to detecting $\rho_{a}$, $\Pi_{a}$, is the projection onto the subspace spanned by the eigenvectors of $\Lambda$ with positive eigenvalues, and the POVM element corresponding to detecting $\rho_{b}$, $\Pi_{b}$, is the projection onto the subspace spanned by the eigenvectors of $\Lambda$ with either negative or zero eigenvalues (the states with eigenvalue zero can be placed in either POVM element, we have chosen to include them in the one corresponding to $\rho_{b}$). 

In our case, we can evaluate the trace norm.  Note that we have
\begin{eqnarray}
\label{lambda1}
\Lambda & = & \| Q_{+}(A_{+}|\psi\rangle\langle\psi | A_{+}^{\dagger} - A_{-}|\psi\rangle\langle\psi |A_{-}^{\dagger}) Q_{+} \nonumber \\
& & + Q_{-}(A_{+}|\psi\rangle\langle\psi |A_{+}^{\dagger} - A_{-}|\psi\rangle\langle\psi |A_{-}^{\dagger}) Q_{-} \|_{1} \nonumber \\
& = & \| Q_{+}(A_{+}|\psi\rangle\langle\psi |A_{+}^{\dagger}- A_{-}|\psi\rangle\langle\psi |A_{-}^{\dagger}) Q_{+}\|_{1} \nonumber \\
& & + \| Q_{-}(A_{+}|\psi\rangle\langle\psi |A_{+}^{\dagger}- A_{-}|\psi\rangle\langle\psi |A_{-}^{\dagger}) Q_{-}\|_{1} .
\end{eqnarray}
The trace norm can be split into two parts, because $Q_{+}$ and $Q_{-}$ have orthogonal supports.  In each of the parts, the problem is reduced to finding the trace norm of a two dimensional matrix.  In the first term, the support of the operator is the subspace spanned by the vectors $Q_{+}P_{+}|\psi\rangle$ and $Q_{+}P_{-}|\psi\rangle$, and for the second term the support lies in the subspace spanned by the vectors $Q_{-}P_{+}|\psi\rangle$ and $Q_{-}P_{-}|\psi\rangle$.  We then find that
\begin{eqnarray}
\label{Lambda}
\|\Lambda\|_{1} & = & \left[ \left( \|Q_{+}A_{+}\psi \|^{2} + \|Q_{+}A_{-}\psi\|^{2}\right)^{2} \right. \nonumber \\
& & \left. - 4|\langle\psi |A_{-}^{\dagger}Q_{+}A_{+}\psi\rangle |^{2}\right]^{1/2} \nonumber \\
& & + \left[ \left( \|Q_{-}A_{+}\psi \|^{2} + \|Q_{-}A_{-}\psi\|^{2}\right)^{2} \right. \nonumber \\
& & \left. - 4|\langle\psi |A_{-}^{\dagger}Q_{-}A_{+}\psi\rangle |^{2}\right]^{1/2} .
\end{eqnarray}
Now $\|\Lambda\|_{1}$ is between $0$ and $1$, with $\|\Lambda\|_{1} =1$ corresponding to perfectly distinguishable states and $\|\Lambda\|_{1} = 0$ corresponding to states that cannot be distinguished.  In our case, if $Q_{+}A_{-}|\psi\rangle = Q_{-}A_{+}|\psi\rangle = 0$, then we will have $\Lambda =1$. In this case, the result of the first measurement determines the result of the second measurement.  In order for $\Lambda =0$, it must be the case that $Q_{+}A_{+}|\psi\rangle = e^{i\phi_{1}}Q_{+}A_{-}|\psi\rangle$ and $Q_{-}A_{+}|\psi\rangle = e^{i\phi_{2}}Q_{-}A_{-}|\psi\rangle$ for some $\phi_{1}$ and $\phi_{2}$.  For the case of a qubit, this can occur when $|\psi\rangle = |0\rangle$, $A_{\pm}=|\pm x\rangle\langle \pm x|$, where $|\pm x\rangle = (|0\rangle \pm |1\rangle )/\sqrt{2}$, and $Q_{+}=|0\rangle\langle 0|$ and $Q_{-}=|1\rangle\langle 1|$.

For qubits, we can go further.  Assuming that $Q_{\pm}$ are rank one projections, the inner products in Eq.\ (\ref{Lambda}) factorize, and we have that
\begin{equation}
\label{qubit-Lambda}
\Lambda = | P(+,+)-P(+,-)| + |P(-,+)-P(-,-)| ,
\end{equation}
where $P(j,k)=\| Q_{j}A_{k}\psi \|^{2}$, for $j,k\in \{ +,-\}$, and $P(j,k)$ is the probability that the first measurement gives the result $k$ and the second gives $j$.  From this, we see that if the probabilities of the different measurement outcomes are close to the same, it will be difficult to distinguish the output states corresponding to different values of the first measurement.  For qudits, the expression on the right-hand side of Eq.\ (\ref{qubit-Lambda}) is a lower bound for $\Lambda$, so its value gives a worst case for ones ability to determine the result of the first measurement.

Now let us look at determining the results of both measurements.  In the case we have been considering so far, in which the second measurement is a projective one is straightforward, because the projections $Q_{+}$ and $Q_{-}$ have orthogonal support, which implies that the states that result from different outcomes for the second measurement are perfectly distinguishable.  This also makes it simple to determine the results of both measurements.  First  we measure the output state in order to determine whether it is in the support of $Q_{+}$ or $Q_{-}$.  That reduces the problem to one of distinguishing between two states, for example, if the output state was found to be in the support of $Q_{+}$, then we would need to discriminate between $Q_{+}P_{+}|\psi\rangle /\| Q_{+}P_{+}\psi\|$ and $Q_{+}P_{-}|\psi\rangle /\| Q_{+}P_{-}\psi\|$.  This can then be accomplished by using minimum error discrimination.  The success probability for determining both measurements using this procedure is the same as that of determining the result of the first measurement, $(1+\|\Lambda\|_{1})/2$, where $\Lambda$ is given by Eq.\ (\ref{Lambda}).  This is shown in greater detail in Appendix A, and it is also shown there that this procedure is optimal. 

In the next section, we will look at the case when both measurements are POVM's for a simple example, a double qubit interferometer.  We will see what one can learn about the path taken through the interferometer, which is specified by the results of two measurements, by measuring the final state of the qubit.  Can one learn more about the path of the qubit if the measurements are less disturbing and, therefore, interfere with each other less?  In particular, if the second measurement is not a projection, one would expect more information about the result of the first measurement to make it through to the final state.  Our model allows us to examine this idea.

\section{Model}
We shall consider a qubit double interferometer based on the qubit single interferometer used by Englert to derive a visibility-path-information duality relation \cite{englert}.  This will allow us to consider measurements other than projective measurements.  We start the qubit in the state $|0\rangle$ and it then passes through a Hadamard gate, which puts it in the state $|+x\rangle = (|0\rangle + |1\rangle )/\sqrt{2} = H|0\rangle$, where we have denoted the operator corresponding to the Hadamard gate by $H$.  Note that $H|1\rangle = |-x\rangle = (|0\rangle - |1\rangle )/\sqrt{2}$.  We then measure which path the qubit took, by which we mean whether it is in the state $|0\rangle$ or $|1\rangle$.  The qubit then passes through a second Hadamard gate, and we again measure whether it is in the state $|0\rangle$ or $|1\rangle$.  The qubit then passes through a final Hadamard gate.  We can view the measurement results as defining a trajectory that the qubit follows through the interferometer, and we are interested in determining what information we can gain about the trajectory by measuring the state of the qubit when it emerges from the interferometer.  

The measurements will not necessarily extract all of the information about the qubit's state so that we can examine the relation between how much path information is extracted and the final state of the qubit.  To measure the qubit going through the interferometer (qubit $a$) we couple it first to a second qubit (qubit $b$), which is initially in the state $|0\rangle_{b}$, using the unitary operation
\begin{eqnarray}
U|0\rangle_{a}|0\rangle_{b} & = & |0\rangle_{a} |\eta (-\theta )\rangle_{b} \nonumber \\
U|1\rangle_{a}|0\rangle_{b} & = & |1\rangle_{a} |\eta (\theta )\rangle_{b} ,
\end{eqnarray}
where $|\eta (\theta )\rangle = \cos\theta |0\rangle + \sin\theta |1\rangle$.  The parameter $0\leq \theta \leq \pi /4$ controls how much information the measurement extracts about the path.  If $\theta =0$ no path information is extracted, while if $\theta = \pi /4$ the maximum amount of information is extracted.  When we measure the auxiliary qubit, we perform the optimal minimum error measurement to distinguish $|\eta (\theta )\rangle_{b}$ and $|\eta (-\theta )\rangle_{b}$ \cite{helstrom}.  That means we measure in the basis $|\pm x\rangle_{b}$.  We shall interpret the result $-$, meaning $|-x\rangle_{b}$ as corresponding to qubit $a$ being in the state $|0\rangle_{a}$ and $+$ corresponding to qubit $a$ being in the state $|1\rangle_{a}$.  Let us now find the measurement operators corresponding to this procedure.  If the pre-measurement state is $|0\rangle_{a}$ and we obtain $|+x\rangle_{b}$ as the measurement result, the post-measurement state is
\begin{equation}
\label{A-state1}
A_{+}|0\rangle_{a} = \,_{b}\langle +x|\eta (-\theta )\rangle_{b} |0\rangle_{a} = \frac{1}{\sqrt{2}}(\cos\theta -\sin\theta ) |0\rangle_{a} .
\end{equation}
Similarly, we find that
\begin{eqnarray}
\label{A-state2}
A_{+}|1\rangle_{a} & = & \frac{1}{\sqrt{2}}( \cos\theta + \sin\theta ) |1\rangle_{a} \nonumber \\
A_{-}|0\rangle_{a} & = & \frac{1}{\sqrt{2}}( \cos\theta + \sin\theta ) |0\rangle_{a} \nonumber \\
A_{-}|1\rangle_{a} & = & \frac{1}{\sqrt{2}}( \cos\theta - \sin\theta ) |1\rangle_{a} .
\end{eqnarray}
In terms of matrices in the $\{ |0\rangle , |1\rangle \}$ basis we have
\begin{eqnarray}
\label{A-matrix}
A_{+} & = & \frac{1}{\sqrt{2}} \left(\begin{array}{cc} \cos\theta - \sin\theta & 0 \\ 0 & \cos\theta + \sin\theta \end{array} \right) \nonumber \\
A_{-} & = & \frac{1}{\sqrt{2}} \left(\begin{array}{cc} \cos\theta + \sin\theta & 0 \\ 0 & \cos\theta - \sin\theta \end{array} \right) .
\end{eqnarray}
The corresponding POVM operators are
\begin{eqnarray}
\Pi_{+} & = & A^{\dagger}_{+}A_{+} = \frac{1}{2}(I -\sin (2\theta )\sigma_{z} ) \nonumber \\
\Pi_{-} & = & A^{\dagger}_{-}A_{-} = \frac{1}{2}(I + \sin (2\theta )\sigma_{z} ) .
\end{eqnarray}

The final states, up to normalization, are given by applying Hadamard operators and the measurement operators to the initial state.  In particular, if both measurements yielded $+x$, then the final state is proportional to $HA_{+}HA_{+}H|0\rangle$ (we shall henceforth drop the subscript $a$ on the qubit).  After the first Hadamard, the state is $|+x\rangle$ and the probabilities of the first measurement are $P(+)=P(-) = 1/2$.  The joint probabilities for the two measurements are given by
\begin{eqnarray}
P(+,+) & = & {\rm Tr}( A_{+}HA_{+}|+x\rangle\langle +x|A^{\dagger}_{+}HA^{\dagger}_{+}) \nonumber \\
 & = & \frac{1}{4}[ 1-\sin (2\theta )\cos (2\theta ) ] .
\end{eqnarray}
Similarly we find
\begin{eqnarray}
\label{probabilities}
P(+,-) & = & P(+,+) =  \frac{1}{4}[ 1-\sin (2\theta )\cos (2\theta ) ]  \nonumber \\
P(-,+) & = & P(-,-) =  \frac{1}{4}[ 1+ \sin (2\theta )\cos (2\theta ) ] ,
\end{eqnarray}
where the first argument in the probability corresponds to the second measurement and the second argument corresponds to the first measurement, i.e.\ $P(+,-)$ is the probability of first getting $-x$ and then getting $+x$ for the measurement results.  This corresponds to the order in which the measurement operators are applied to the state.  The resulting normalized output states, with the same convention for the ordering of the measurement results, are
\begin{eqnarray}
\label{output-states}
|\psi_{out}^{++}\rangle & = & \frac{1}{[ 1-\sin (2\theta )\cos (2\theta )]^{1/2}} [\cos\theta (\cos\theta - \sin\theta ) |+x\rangle \nonumber \\
& & - \sin\theta (\sin\theta +\cos\theta ) |-x\rangle ] \nonumber \\
|\psi_{out}^{+-}\rangle & = & \frac{1}{[ 1-\sin (2\theta )\cos (2\theta )]^{1/2}} [\cos\theta (\cos\theta - \sin\theta ) |+x\rangle \nonumber \\
& & + \sin\theta (\sin\theta +\cos\theta ) |-x\rangle ] \nonumber \\
|\psi_{out}^{-+}\rangle & = & \frac{1}{[ 1+\sin (2\theta )\cos (2\theta )]^{1/2}} [\cos\theta (\cos\theta + \sin\theta ) |+x\rangle \nonumber \\
& & + \sin\theta (\sin\theta - \cos\theta ) |-x\rangle ] \nonumber \\
|\psi_{out}^{--}\rangle & = & \frac{1}{[ 1+\sin (2\theta )\cos (2\theta )]^{1/2}} [\cos\theta (\cos\theta + \sin\theta ) |+x\rangle \nonumber \\
& & - \sin\theta (\sin\theta -\cos\theta ) |-x\rangle ] \nonumber \\
\end{eqnarray}
Note that when $\theta =0$, in which case the measurement extracts no path information, all of these vectors become $|+x\rangle$, and there is no correlation between the final state and the measurement results.  When $\theta =\pi /4$, then $|\psi_{out}^{++}\rangle$ and $|\psi_{out}^{+-}\rangle$ are parallel to $|-x\rangle$ and $|\psi_{out}^{-+}\rangle$ and $|\psi_{out}^{--}\rangle$ are parallel to $|+x\rangle$.  Then we can only distinguish between the two sets, $\{ |\psi_{out}^{++}\rangle , |\psi_{out}^{+-}\rangle\}$ and $\{ |\psi_{out}^{-+}\rangle , |\psi_{out}^{--}\rangle\}$.

If we represent the four output states as vectors in the $\{ |+x\rangle , |-x\rangle \}$ plane, with $|+x\rangle$ being the horizontal direction and $|-x\rangle$ the vertical, we find the following.  The states $|\psi_{out}^{++}\rangle$ and $|\psi_{out}^{+-}\rangle$ make an angle of $-\phi_{2}$ and $\phi_{2}$, respectively, with the horizontal axis, where
\begin{equation}
\tan \phi_{2}=\frac{\cos\theta + \sin\theta}{\cos\theta - \sin\theta} \tan\theta ,
\end{equation}
and $|\psi_{out}^{--}\rangle$ and $|\psi_{out}^{-+}\rangle$ make angles of $\phi_{1}$ and $-\phi_{1}$, respective, with the horizontal axis, where 
\begin{equation}
\tan \phi_{1}=\frac{\cos\theta - \sin\theta}{\cos\theta + \sin\theta} \tan\theta .
\end{equation}
Note that for $0\leq \theta \leq \pi /4$, we have that $\phi_{2}\geq \phi_{1}$ and that $\phi_{2}$ is an increasing function of $\theta$, which goes from $0$ at $\theta =0$ to $\pi /2$ at $\theta = \pi /4$.  The behavior of $\phi_{1}$ is a bit more complicated.  It is $0$ at $\theta =0$, increases and then decreases again becoming $0$ at $\theta = \pi /4$.  Both $\phi_{1}$ and $\phi_{2}$ are plotted as functions of $\theta$ in Fig.\ 1.
\begin{figure}
\label{phi12}
\includegraphics[scale=.40]{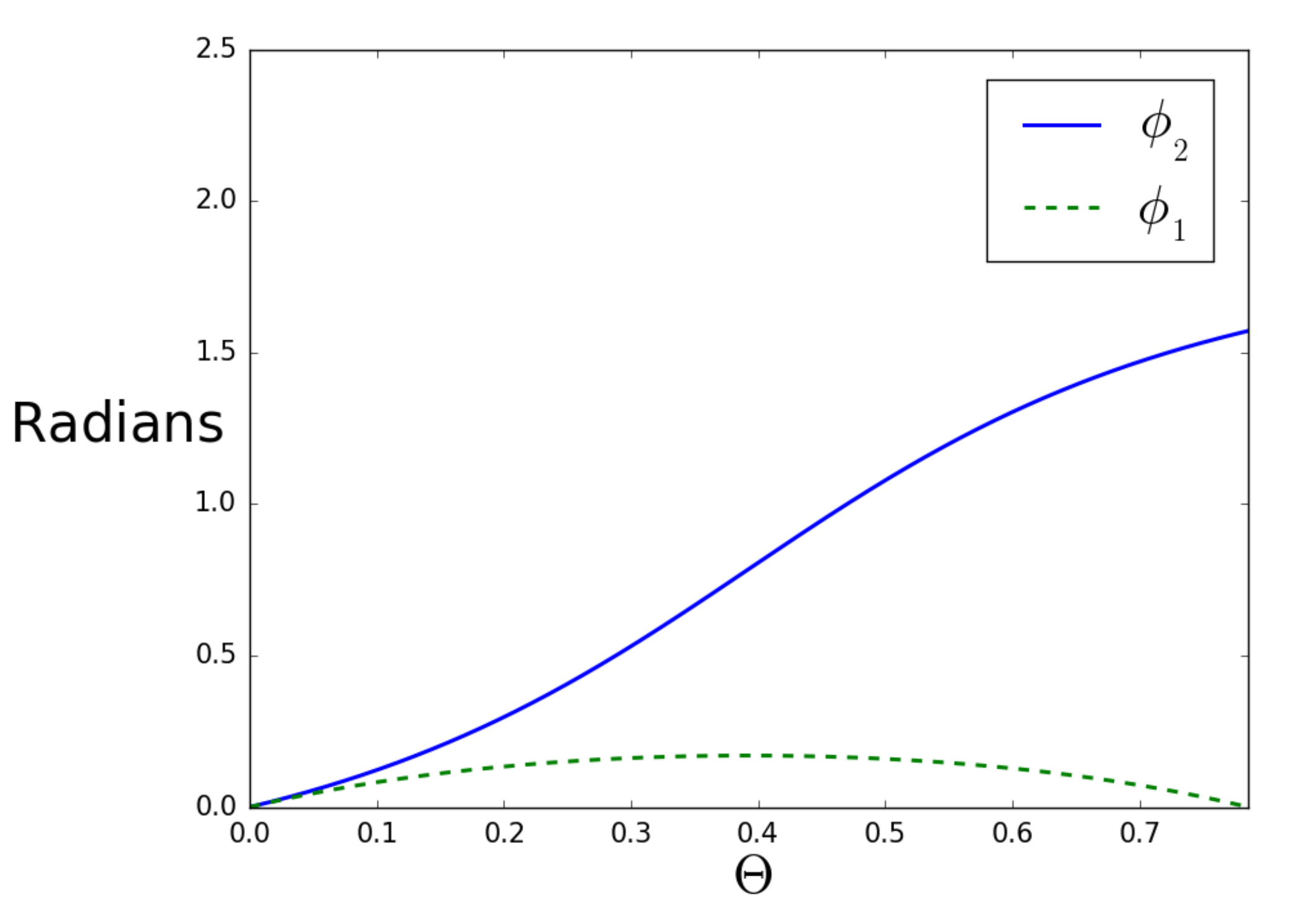}
\caption{The angles $\phi_{1}$ and $\phi_{2}$ plotted as functions of $\theta$.  The dotted line is for $\phi_{1}$ and the solid line is for $\phi_{2}$.} 
\end{figure}

 \section{Measurement of the final state}
 Now that we have the final states, we can ask what kind of information we can learn by measuring them.  There are a number of possibilities.  One is to determine, as best we can, the results of either the first or the second measurement.  Another possibility is to perform a four-outcome measurement that maximizes our probability of finding both measurement results.  A final possibility is to perform a measurement that eliminates some of the possible trajectories.  We shall look at each of these possibilities in turn.
 \subsection{Results of individual measurements}
 Suppose we only wish to determine the result of the second measurement.  The density matrix corresponding to the result $+$, if we ignore the result of the first measurement, is
 \begin{eqnarray}
 \rho_{2+}  & = & \frac{1}{[P(+,+)+P(+,-)]} [P(+,+) |\psi_{out}^{++}\rangle\langle \psi_{out}^{++}| \nonumber \\
 & & + P(+,-)|\psi_{out}^{+-}\rangle\langle \psi_{out}^{+-}| ] .
\end{eqnarray}
and the density matrix corresponding to $-$ is
\begin{eqnarray}
\rho_{2-} & = &  \frac{1}{[P(-,+)+P(-,-)]} [P(-,+) |\psi_{out}^{-+}\rangle\langle \psi_{out}^{-+}| \nonumber \\
 & & + P(-,-)|\psi_{out}^{--}\rangle\langle \psi_{out}^{--}| ] .
\end{eqnarray}
Our problem in determining the result of the second measurement is reduced to discriminating between these two density matrices.  This can be done using minimum-error state discrimination \cite{helstrom}.    In this case, choosing $\rho_{2-}$ as $\rho_{b}$, which occurs with a probability of $P(-,+)+P(-,-)$, and $\rho_{2+}$as $\rho_{a}$, which occurs with a probability of $P(+,+)+P(+,-)$, (see Eq.\ (\ref{min-err})) we find that 
\begin{eqnarray}
\Lambda & = & P(+,+) |\psi_{out}^{++}\rangle\langle \psi_{out}^{++}| + P(+,-)|\psi_{out}^{+-}\rangle\langle \psi_{out}^{+-}| \nonumber \\
& & - P(-,+) |\psi_{out}^{-+}\rangle\langle \psi_{out}^{-+}| - P(-,-)|\psi_{out}^{--}\rangle\langle \psi_{out}^{--}| , \nonumber \\
& = & -2\sin\theta \cos^{3}\theta |+x\rangle\langle +x|  \nonumber \\  
& & + 2\sin^{3}\theta \cos\theta |-x\rangle\langle -x| .
\end{eqnarray}
This implies that the POVM element corresponding to detecting $\rho_{2+}$ is $\Pi_{2+}^{(out)}=|-x\rangle\langle -x|$, and the POVM element corresponding to $\rho_{2-}$ is $\Pi_{2-}^{(out)}=|+x\rangle\langle +x|$.  We shall denote the results of the output measurement as $M_{out}=+$, corresponding to the detection of $\rho_{2+}$, and $M_{out}=-$, corresponding to the detection of $\rho_{2-}$.  Finding the trace norm of $\Lambda$ now gives us that
\begin{equation}
P_{s}=\frac{1}{2}[1+ |\sin (2\theta )|] .
\end{equation}
This result is not surprising in that the density matrices become more distinguishable as $\theta$ goes from $0$ to $\pi /4$.  At $\theta =0$ they are identical and equally probable, so guessing is the best we can do.  At $\theta = \pi /4$ they are also equally probable, but they are now orthogonal and, therefore, perfectly distinguishable.

Using Bayes' theorem we can see what is the effect of updating the probabilities for the occurrence of the two density matrices, $\rho_{2+}$ and $\rho_{2-}$, which is also the same as updating the probabilities for the result of the second measurement.  Let us denote by $P(\rho_{2+})$ and $P(\rho_{2-})$ the probabilities of obtaining $\rho_{2+}$ and $\rho_{2-}$ at the output, respectively.  These are given by
\begin{eqnarray}
\label{bef-meas2}
P(\rho_{2+}) & = & P(+,+) + P(+,-) = \frac{1}{2}[ 1-\sin (2\theta )\cos (2\theta ) ] \nonumber \\
P(\rho_{2-}) & = & P(-,+) + P(-,-) = \frac{1}{2}[ 1+ \sin (2\theta )\cos (2\theta ) ] . \nonumber \\
\end{eqnarray}
What we would like to find are the probabilities of the occurrence of $\rho_{2+}$ and $\rho_{2-}$ conditioned on the result of the measurement on the output state.  This is the same as finding the probabilities of the result of the second measurement conditioned on the measurement of the output state.  We shall denote these probabilities by $P(\rho_{2j}|M_{out}=k)$, where, $j,k = \pm$, i.e.\ the probability of $\rho_{2j}$ occurring if the measurement of the output state is $M_{out}=k$.  Bayes' theorem tells us that
\begin{equation}
P(\rho_{2j}|M_{out}=k)P(M_{out}=k)=P(M_{out}=k|\rho_{2j})P(\rho_{2j}) ,
\end{equation}
where $P(M_{out}=k|\rho_{2j})= {\rm Tr}(\Pi_{2k}\rho_{2j})$, and
\begin{eqnarray}
P(M_{out}=k) & = & P(M_{out}=k|\rho_{2+}) P(\rho_{2+}) \nonumber \\
& & +P(M_{out}=k|\rho_{2-}) P(\rho_{2-}) .
\end{eqnarray}
From this we find that
\begin{eqnarray}
\label{aft-meas2}
P(\rho_{2+}|M_{out}=+) & = & \frac{1}{2} [ 1+ \sin (2\theta )] \nonumber \\
P(\rho_{2-}|M_{out}=+) & = & \frac{1}{2} [ 1- \sin (2\theta )] \nonumber \\
P(\rho_{2+}|M_{out}=-) & = & \frac{1}{2} [ 1 - \sin (2\theta )] \nonumber \\
P(\rho_{2-}|M_{out}=-) & = & \frac{1}{2} [ 1+ \sin (2\theta )] .
\end{eqnarray}
Now suppose we measured the output state and obtained $M_{out}=+$.  Before the measurement the probability of the output state being being $\rho_{2+}$ was $P(\rho_{2+})$, while after the measurement it is $P(\rho_{2+}|M_{out}=+)$, and a comparison of Eqs.~(\ref{bef-meas2},\ref{aft-meas2}) shows that $P(\rho_{2+}|M_{out}=+)\geq P(\rho_{2+})$.  Therefore, the result of the measurement on the output state has increased the probability that the result of the second measurement was indeed $+$, and the difference between $P(\rho_{2+}|+)$ and $P(\rho_{2+})$ is an increasing function of $\theta$.

The situation becomes more interesting if we wish to determine only the result of the first measurement.  Ignoring the result of the second measurement, the output density matrix corresponding to the result $+$ for the first measurement is
\begin{eqnarray}
\rho_{1+} & = & \frac{1}{[P(+,+)+P(-,+)]} [ P(+,+) |\psi_{out}^{++}\rangle\langle \psi_{out}^{++}| \nonumber \\
& & + P(-,+) |\psi_{out}^{-+}\rangle\langle \psi_{out}^{-+}| ] ,
\end{eqnarray}
and the output density matrix corresponding to $-$ is
\begin{eqnarray}
\rho_{1-} & = & \frac{1}{[P(+,-)+P(-,-)]} [ P(+,-)|\psi_{out}^{+-}\rangle\langle \psi_{out}^{+-}| \nonumber \\
& & + P(-,-)|\psi_{out}^{--}\rangle\langle \psi_{out}^{--}| ] ,
\end{eqnarray}
and the probabilities of these output density matrices occurring are $P(\rho_{1+})=P(\rho_{1-})=1/2$.  We can now find the optimal minimum-error discrimination for this situation, and we find that the POVM elements are $\Pi_{1+}=|1\rangle\langle 1|$ and $\Pi_{1-}=|0\rangle\langle 0|$, and the success probability is now
\begin{equation}
P_{s}=\frac{1}{2} + \frac{1}{4}\sin (4\theta ) .
\end{equation}
This has a different behavior than the success probability for the second measurement.  It is $1/2$ at $\theta = 0$ since, again, the states are identical and equally probable, and then increases reaching a maximum value of $3/4$ at $\theta = \pi /8$.  It then decreases back to $1/2$ at $\theta = \pi /4$.  The reason for the decrease is that the states are equally probable for the entire range of $\theta$, and as $\theta$ approaches $\pi /4$, the second measurement becomes closer to a projective measurement, and this eliminates the correlation between the first measurement and the final state.

The success probability in discriminating the two output states resulting from either of the measurements (in this case the first or the second) serves as a useful measure or the influence of the measurement on the output state.  In the case of the first measurement, the influence for $\theta$ small is small, then grows, but subsequently declines as the second measurement forces $\rho_{1+}$ and $\rho_{1-}$ to become less distinguishable.

As before, we can use Bayes' theorem to find the probabilities of $\rho_{1\pm}$, that is the probabilities of the results of the first measurement, conditioned on a result of the measurement of the output state.  We now let the output state measurement result $M_{out}=+$ correspond to $\Pi_{1+}$ and $M_{out}=-$ correspond to $\Pi_{1-}$.  In analogy with what we did before, we find that
\begin{eqnarray}
\label{aft-meas1}
P(\rho_{1+}|M_{out}=+) & = & \frac{1}{2} [ 1+ \sin (2\theta )\cos (2\theta )] \nonumber \\
P(\rho_{1-}|M_{out}=+) & = & \frac{1}{2} [ 1- \sin (2\theta )\cos (2\theta )] \nonumber \\
P(\rho_{1+}|M_{out}=-) & = & \frac{1}{2} [ 1 - \sin (2\theta )\cos (2\theta )] \nonumber \\
P(\rho_{1-}|M_{out}=-) & = & \frac{1}{2} [ 1+ \sin (2\theta )\cos (2\theta )] .
\end{eqnarray}
Note that in this case, the difference between, for example, $P(\rho_{1+}|M_{out}=+)$ and $P(\rho_{1+})=1/2$ first increases with $\theta$ as the measurements extract more information, but then decreases as the second measurement interferes with the first.  

\subsection{Retrodiction of the trajectory}
Instead of trying to determine the result of either the first or second measurement, one can try to determine both.  We then need a measurement that will discriminate among the four output states in Eq.\ (\ref{output-states}).  Unfortunately an explicit form for the optimal minimum-error measurement is only known for two states, so we will have to proceed in a different manner than we have so far.  First, we will use a pretty good discrimination measurement, the square-root measurement \cite{croke}.  Next we will numerically find an optimal discrimination measurement, and compare its success probability to that of the square-root measurement.

Suppose we want to discriminate among the states  $\{ |\psi_{j}\rangle |j=1,2,\ldots N\}$, where $\psi_{j}$ occurs with probability $p_{j}>0$.  The POVM elements for the square root measurement are given by 
\begin{equation}
\Pi_{j} = p_{j}\rho^{-1/2}|\psi_{j}\rangle\langle \psi_{j}|\rho^{-1/2} ,
\end{equation}
where $\rho = \sum_{j=1}^{N}p_{j} |\psi_{j}\rangle\langle \psi_{j}|$, and the inverse is take on the span of the vectors $\{ |\psi_{j}\rangle |j=1,2,\ldots N\}$.  In our case we find that
\begin{equation}
\rho = \cos^{2}\theta |+x\rangle\langle +x| + \sin^{2}\theta |-x\rangle\langle -x| ,
\end{equation}
so that 
\begin{equation}
\rho^{-1/2}= \frac{1}{\cos\theta}|+x\rangle\langle +x| + \frac{1}{\sin\theta} |-x\rangle\langle -x| .
\end{equation}
defining the states
\begin{eqnarray}
|\tilde{\psi}^{++} \rangle & = & (\cos\theta -\sin\theta )|+x\rangle - (\sin\theta + \cos\theta )|-x\rangle \nonumber \\
|\tilde{\psi}^{+-}\rangle & = & (\cos\theta -\sin\theta )|+x\rangle + (\sin\theta + \cos\theta )|-x\rangle \nonumber \\
|\tilde{\psi}^{-+}\rangle & = & (\cos\theta +\sin\theta )|+x\rangle + (\sin\theta - \cos\theta )|-x\rangle
\nonumber \\
|\tilde{\psi}^{--}\rangle & = & (\cos\theta +\sin\theta )|+x\rangle - (\sin\theta - \cos\theta )|-x\rangle ,
\nonumber \\
\end{eqnarray}
the POVM elements for the square-root measurement are 
\begin{equation}
\Pi_{jk}= \frac{1}{4} |\tilde{\psi}^{jk}\rangle\langle \tilde{\psi}^{jk}| ,
\end{equation}
where $j,k = \pm$.  The probability of successfully identifying the state is
\begin{eqnarray}
P_{s} & = & \sum_{j,k=\pm} P(j,k)\langle \psi_{out}^{jk}|\Pi_{jk}|\psi_{out}^{jk}\rangle \nonumber \\
 & = & \frac{1}{4} [ 1 + \sin (2\theta )+ \sin^{2}(2\theta ) -\sin^{3}(2\theta )] .
\end{eqnarray}

It is useful to compare this to the optimal minimum-error measurement for these states, which we shall find numerically.  The set of four states we are trying to discriminate is invariant under a reflection about the $|+x\rangle$ axis, so the POVM elements should also have this property \cite{andersson}.  Consequently, we choose $\Pi_{--}=c_{1}|\xi_{1}\rangle\langle\xi_{1}|$, $\Pi_{+-}=c_{2}|\xi_{2}\rangle\langle\xi_{2}|$, $\Pi_{-+}=c_{1}|\xi_{3}\rangle\langle\xi_{3}|$, and $\Pi_{++}= c_{2}|\xi_{4}\rangle\langle\xi_{4}|$, where
\begin{eqnarray}
|\xi_{1}\rangle & = & \cos\mu_{1} |+x\rangle + \sin\mu_{1}|-x\rangle \nonumber \\
|\xi_{2}\rangle & = & \cos\mu_{2} |+x\rangle + \sin\mu_{2} |-x\rangle ,
\end{eqnarray}
$|\xi_{3}\rangle$ is just  $|\xi_{1}\rangle$ with $\mu_{1}$ replaced by $-\mu_{1}$ and $|\xi_{4}\rangle$ is just $|\xi_{2}\rangle$ with $\mu_{2}$ replaced by $-\mu_{2}$.  We also have that $c_{1}$ and $c_{2}$ are between $0$ and $1$.  The requirement that the POVM elements sum to the identity gives us that
\begin{eqnarray}
\label{id-cond}
c_{1}\cos^{2}\mu_{1} + c_{2}\cos^{2}\mu_{2} & = & \frac{1}{2} \nonumber \\
c_{1}\sin^{2}\mu_{1} + c_{2} \sin^{2}\mu_{2} & = & \frac{1}{2} .
\end{eqnarray}
Adding these equations we find that $c_{1}+c_{2}=1$ and subtracting them gives $c_{1}\cos (2\mu_{1})+c_{2}\cos (2\mu_{2}) = 0$.  These equations will have a solution in the range $0\leq c_{1},c_{2}\leq 1$ if either $1/2\geq \sin^{2}\mu_{1}$ and $1/2\leq \sin^{2}\mu_{2}$, or $1/2\leq \sin^{2}\mu_{1}$ and $1/2\geq \sin^{2}\mu_{2}$.  We will choose $0\leq \mu_{1} \leq \pi /4$ and $ \pi /4 \leq \mu_{2} \leq \pi /2$, which guarantees that this condition is satisified.  Solving these equations for $c_{1}$ and $c_{2}$, we find that
\begin{eqnarray}
c_{1} & = & \frac{-\cos (2\mu_{2})}{\cos (2\mu_{1}) - \cos (2\mu_{2})} \nonumber \\
c_{2} & = & \frac{\cos (2\mu_{1})}{\cos (2\mu_{1}) - \cos (2\mu_{2})} .
\end{eqnarray}

In our case, the states $\psi_{out}^{-+}$ and $\psi_{out}^{--}$ occur with a probability $p_{1}= P(-,-)$ and $\psi_{out}^{++}$ and $\psi_{out}^{+-}$ occur with a probability of $p_{2}=P(+,+)$, where $p_{1}+p_{2}=1/2$.  The success probability is now
\begin{equation}  
P_{s}=2 \frac{-p_{1}\cos (2\mu_{2})\cos (\mu_{1}-\phi_{1}) +p_{2}\cos (2\mu_{1}) \cos (\mu_{2} -\phi_{2})}{\cos (2\mu_{1}) - \cos (2\mu_{2})}
\end{equation}
For each value of $\theta$, which determines the values of $\phi_{1}$ and $\phi_{2}$, we can do a search in the allowed ranges of $\mu_{1}$ and $\mu_{2}$ in order to find values that maximize the above expression.  The results are shown in Fig.\ 2. These results are surprising.  Both the square-root measurement and the numerical results show that the success probability is greatest at $\theta =\pi /4$, where we can determine with certainty the result of the second measurement, but lose all information about the first.  One might have thought that an intermediate value of $\theta$ would give the greatest value, because in that case the final state would depend on the results of both measurements.  As one can see, however, that is not the case.
\begin{figure}
\label{double-inf}
\includegraphics[scale=.30]{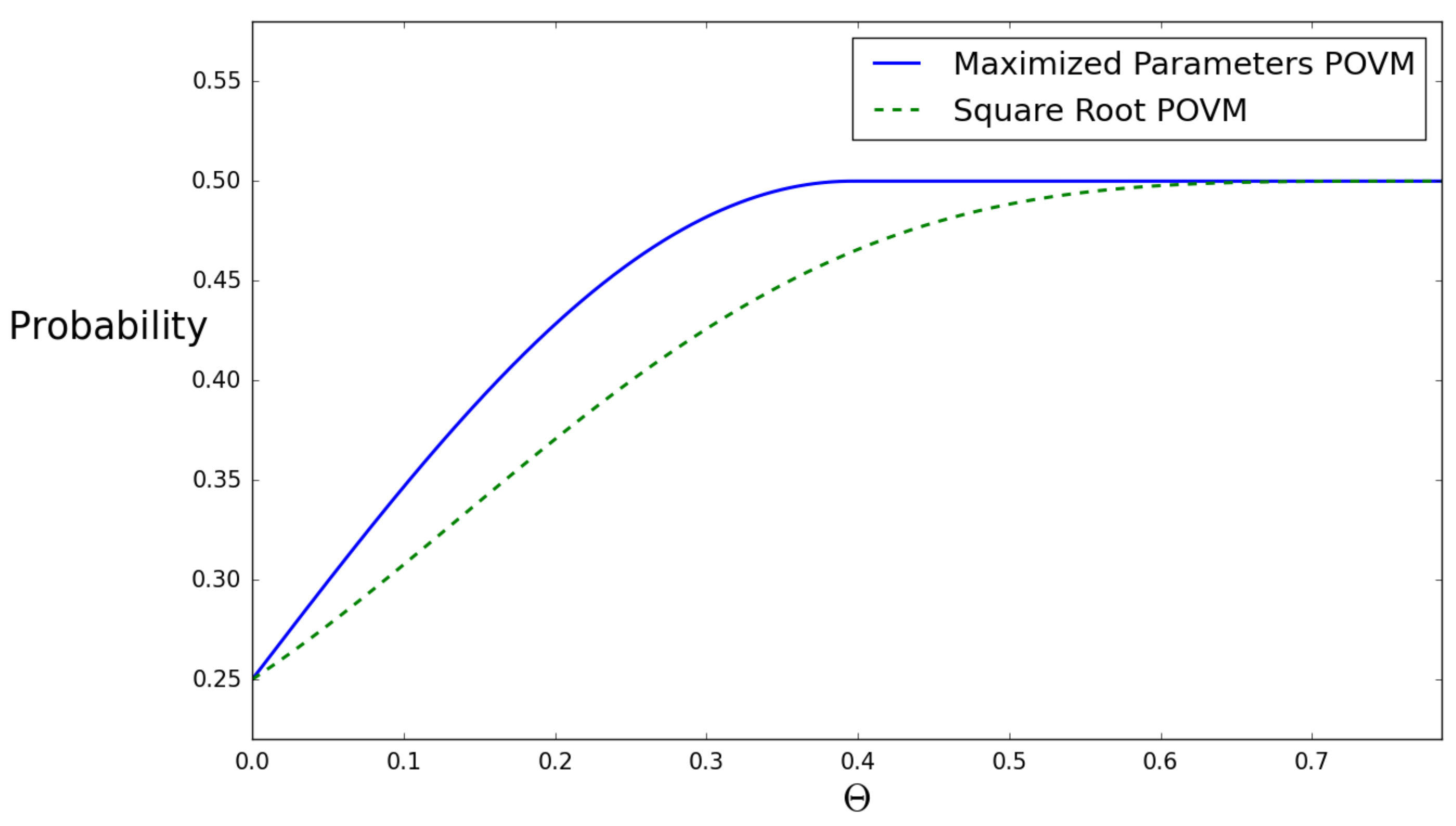}
\caption{Success probability for finding the trajectory versus $\theta$ for a two-loop interferometer.  The dashed line corresponds to the square-root measurement, and the solid line to the numerically optimized measurement.} 
\end{figure}

\subsection{Eliminating a trajectory}
So far, all of the information we have gained about possible trajectories is probabilistic, we can identify likely trajectories, but we cannot say that one definitely occurred.  Is there a measurement we can make that will allow us to say something definite about a trajectory?  The answer to this question is yes if instead of asking which trajectory occurred, we ask if there is one that did not occur.

Measurements can be used to identify states, but they can also be used to eliminate states from a known set \cite{barnett-book,jain}.  This type of measurement has proven useful in quantum digital signature schemes \cite{collins}.  Here we would like to develop a measurement that eliminates one of the four possible trajectories.  Each POVM element will be a projection onto a vector orthogonal to one of the four output states, that is, when acting on one of the output states the result is zero.  If we obtain the measurement result corresponding to that POVM element, then the output state cannot be the state that is annihilated by that element.

Since the set of states we are considering is invariant under a reflection about the $|+x\rangle$ axis, we can construct the POVM elements from the vectors $|\xi_{j}\rangle$, for $j=1,2,3,4$ from the previous section.  We again choose, $\Pi_{--}=c_{1}|\xi_{1}\rangle\langle\xi_{1}|$, $\Pi_{+-}=c_{2}|\xi_{2}\rangle\langle\xi_{2}|$, $\Pi_{-+}=c_{1}|\xi_{3}\rangle\langle\xi_{3}|$, and $\Pi_{++}= c_{2}|\xi_{4}\rangle\langle\xi_{4}|$.  The conditions that guarantee that the POVM elements sum to the identity are given in Eq.\ (\ref{id-cond}).

We will consider $\theta$ in the range $\pi /8 \leq \theta \leq \pi /4$, which implies that $\phi_{1}$ is between $0$ and $\pi /4$, and $\phi_{2}$ is between $\pi /4$ and $\pi /2$.  Define the vectors
\begin{eqnarray}
|\gamma^{++}\rangle & = & \cos (\phi_{2}+\pi /2)|+x\rangle - \sin (\phi_{2}+\pi /2)|-x\rangle \nonumber \\
|\gamma^{+-}\rangle & = & \cos (\phi_{2}+\pi /2)|+x\rangle + \sin (\phi_{2}+\pi /2)|-x\rangle \nonumber \\
|\gamma^{-+}\rangle & = & \cos (\phi_{1}+\pi /2)|+x\rangle - \sin (\phi_{1}+\pi /2)|-x\rangle \nonumber \\
|\gamma^{--}\rangle & = & \cos (\phi_{1}+\pi /2)|+x\rangle + \sin (\phi_{1}+\pi /2)|-x\rangle . \nonumber \\
\end{eqnarray}
We then have the relation $\langle \gamma^{jk}|\psi^{jk}\rangle = 0$ for $j,k=\pm$.  Now we can set $\mu_{1}=\phi_{1}+\pi /2$ and $\mu_{2} = \phi_{2}+\pi /2$, which leads to the identification $\xi_{1}\leftrightarrow \gamma^{--}$, $\xi_{2}\leftrightarrow \gamma^{+-}$, $\xi_{3}\leftrightarrow \gamma^{-+}$ and $\xi_{4}\leftrightarrow \gamma^{++}$.  We have that $\sin^{2}(\phi_{1}+\pi /2) \geq 1/2$ and $\sin^{2}(\phi_{2}+\pi /2) \leq 1/2$, so the conditions for the POVM elements to sum to the identity are fulfilled.  The POVM elements are
\begin{eqnarray}
\Pi_{--} = c_{1}|\gamma^{--}\rangle\langle \gamma^{--}| & \hspace{5mm} & \Pi_{-+}=c_{1}|\gamma^{-+}\rangle\langle \gamma^{+-}| \nonumber \\
\Pi_{+-} =  c_{2}|\gamma^{-+}\rangle\langle \gamma^{-+}| & \hspace{5mm} & \Pi_{++} =  c_{2}|\gamma^{--}\rangle\langle \gamma^{--}| , \nonumber \\
\end{eqnarray}
where 
\begin{equation}
c_{1} = \frac{1-2\cos^{2}\phi_{2}}{2(\cos^{2}\phi_{1}-\cos^{2}\phi_{2})} \hspace{5mm} c_{2}=\frac{1-2\cos^{2}\phi_{1}}{2(\cos^{2}\phi_{2} - \cos^{2}\phi_{1})} .
\end{equation}
If we measure the states with this POVM and obtain the result corresponding to $\Pi_{jk}$, where $j,k=\pm$, then that means the output state was not $|\psi^{jk}\rangle$.

This type of measurement can be used to generate a guess for the trajectory that is guaranteed to have at least one of the measurement results correct.  If the party measuring the final state obtains the result corresponding to $|\gamma^{jk}\rangle$, which means that the trajectory $(k,j)$ did not occur ($k$ is the result of the first measurement, $j$ the result of the second), then the guess for the trajectory should be $(\bar{k},\bar{j})$, where the bar indicates taking the opposite sign, e.g.\ if $j=+$, then $\bar{j}=-$.  To see how this works, suppose we find that the trajectory $(+,+)$ did not occur, so we guess $(-,-)$.  Now since $(+,+)$ did not occur, the possibilities are $(+,-)$, $(-,+)$, and $(-,-)$.  The guess, $(-,-)$ matches the first two possibilities in one place and matches the third possibility completely.  A similar situation arises when trying to find the state of two qubits each of which is in one of two nonorthogonal states (see \cite{wallden}).

\section{ Triple Inferometer }
It is useful to extend the interferometer from two loops to three in order to see how our ability to retrodict trajectories changes as the trajectories become longer.  We will explore a measurement derived from the square-root measurement and one derived numerically.  In this case, we have eight, instead of four, possible output states.  These states are derived from the ones in Eq.\ (\ref{output-states}) by applying either $HA_{+}$ or $HA_{-}$ to them.  Non-normalized versions of these states are given in the appendix. In particular, the state $|\tilde{\psi}^{jkl}\rangle$, where $j,k,l\in \{ +,-\}$, is given by 
\begin{equation}
|\tilde{\psi}^{jkl}\rangle = HA_{j}HA_{k}HA_{l}|+x\rangle ,
\end{equation}
with explicit expressions given in Eq.\ (\ref{3-loop-out}).  

The density matrix, $\rho$, that appears in the square root measurement is, in this case, 
\begin{eqnarray}
\rho & = & \frac{1}{2}[ |+x\rangle\langle +x| + |-x\rangle\langle -x| \nonumber \\
& & + \cos^{2}(2\theta ) (|+x\rangle\langle -x|+|-x\rangle\langle +x|)] ,
\end{eqnarray} 
so that 
\begin{equation}
\rho^{-1/2}= \frac{1}{\sqrt{1+\cos^{2}(2\theta )}} |0\rangle\langle 0| + \frac{1}{\sqrt{1-\cos^{2}(2\theta )}} |1\rangle\langle 1| .
\end{equation}
The POVM elements are given by
\begin{equation}
\Pi_{jkl}=\rho^{-1/2}|\tilde{\psi}^{jkl}\rangle\langle\tilde{\psi}^{jkl}| \rho^{-1/2} .
\end{equation}
Once one has the POVM, calculation of the success probability of the measurement, $P_{s}$, is straightforward, and a plot of $P_{s}$ versus $\theta$ is give for the three-loop case in Fig.\ 3.

\begin{figure}
\label{triple-inf}
\includegraphics[scale=.35]{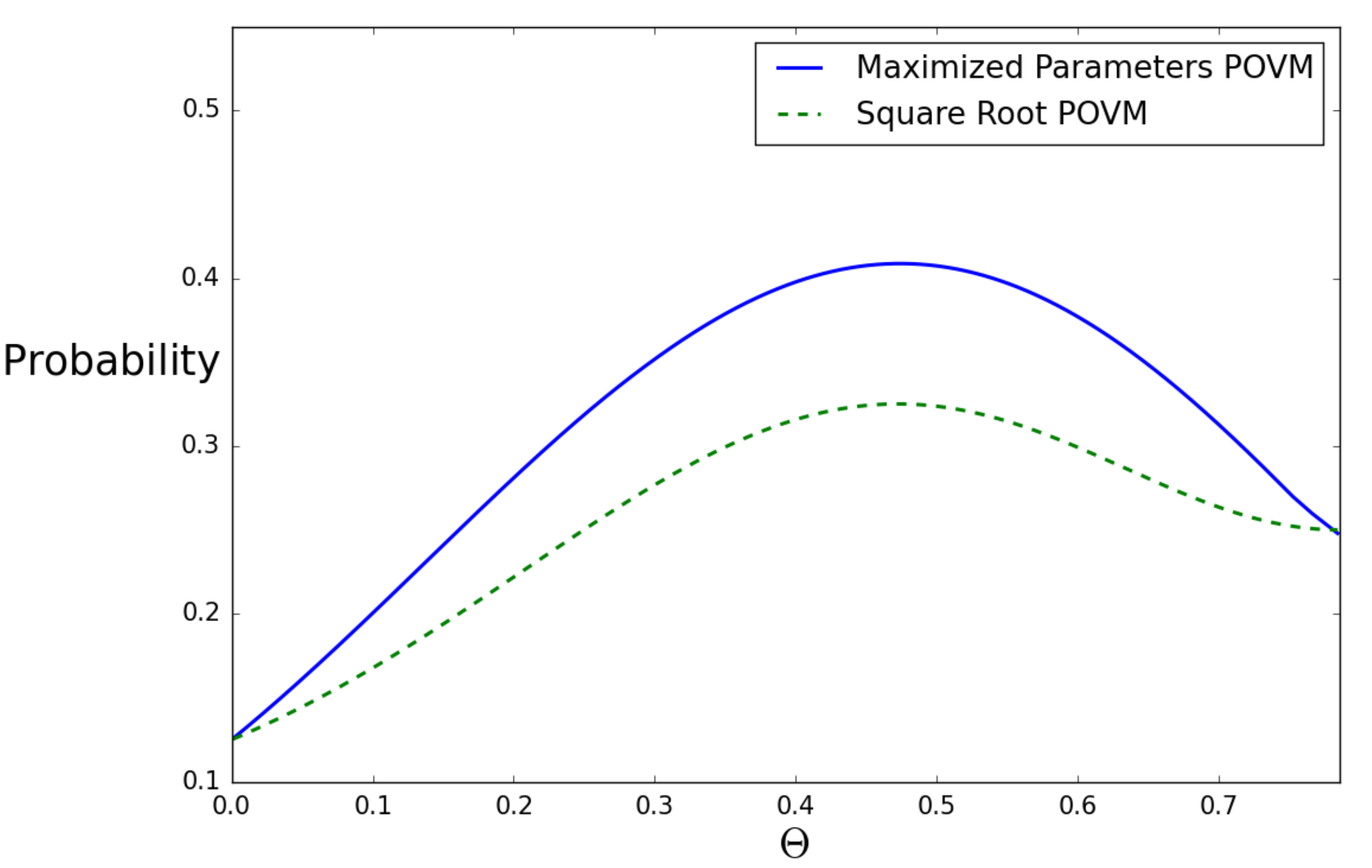}
\caption{Success probability for finding the trajectory versus $\theta$ for a three-loop interferometer.  The dashed line corresponds to the square-root measurement, and the solid line to the numerically optimized measurement.} 
\end{figure}

 We also used a numerical approach to optimize the POVM.  In this case we note that the set of output states is invariant under reflections about the state $|0\rangle$, for example, $|\tilde{\psi}^{+++}\rangle$ and $|\tilde{\psi}^{-+-}\rangle$ are taken into each other by this reflection.  Therefore, our POVM elements will also have this symmetry, so we have
\begin{eqnarray}
\Pi_{+++} = c_{1}|\xi_{1}\rangle\langle \xi_{1}| & \hspace{5mm} & \Pi_{-+-} = c_{8}|\xi_{8}\rangle\langle \xi_{8}| \nonumber \\
\Pi_{+-+} = c_{2}|\xi_{2}\rangle\langle \xi_{2}| & \hspace{5mm} & \Pi_{---} = c_{7}|\xi_{7}\rangle\langle \xi_{7}| \nonumber \\
\Pi_{-++} = c_{3}|\xi_{3}\rangle\langle \xi_{3}| & \hspace{5mm} & \Pi_{++-} = c_{6}|\xi_{6}\rangle\langle \xi_{6}| \nonumber \\
\Pi_{+--} = c_{4}|\xi_{4}\rangle\langle \xi_{4}| & \hspace{5mm} & \Pi_{--+} = c_{5}|\xi_{5}\rangle\langle \xi_{5}| \nonumber \\
\end{eqnarray}
where
\begin{eqnarray}
|\xi_{1}\rangle & = & \cos\mu_{1} |0\rangle + \sin\mu_{1}|1\rangle \nonumber \\
|\xi_{2}\rangle & = & \cos\mu_{2} |0\rangle + \sin\mu_{2} |1\rangle \nonumber\\
|\xi_{3}\rangle & = & \cos\mu_{3} |0\rangle + \sin\mu_{3}|1\rangle \nonumber \\
|\xi_{4}\rangle & = & \cos\mu_{4} |0\rangle + \sin\mu_{4} |1\rangle
\end{eqnarray}
$|\xi_{8}\rangle$, $|\xi_{7}\rangle$, $|\xi_{6}\rangle$, and $|\xi_{5}\rangle$ correspond to $|\xi_{1}\rangle$, $|\xi_{2}\rangle$, $|\xi_{3}\rangle$, and $|\xi_{4}\rangle$ respectively, with all the of $\mu$'s going to -$\mu$'s.  It is important to note here that in the previous case, our condition that POVMs sum to Identity reduced our number of free parameters from 4 to 2, leaving only $\mu_{1}$ and $\mu_{2}$. Here, the same condition reduces them from 8 to 6, requiring that in addition to the 4 $\mu$'s we must have 2 of the $c_{j}$'s be free parameters as well.  Choosing to eliminate $c_{4}$ and $c_{3}$, we find
\begin{eqnarray}
c_{3} & = & \frac{1}{\cos(2\mu_{3}) - \cos(2\mu_{4})}\{ c_{1}[\cos(2\mu_{4}) - \cos(2\mu_{1})]  \nonumber \\
& & + c_{2}[\cos(2\mu_{4}) - \cos(2\mu_{2})] - \cos(2\mu_{4})\}  \nonumber \\
c_{4} & = & \frac{1}{\cos(2\mu_{3}) - \cos(2\mu_{4})} \{-c_{1}[\cos(2\mu_{3}) - \cos(2\mu_{1})] \nonumber \\
& & - c_{2}[\cos(2\mu_{3}) - \cos(2\mu_{2})] - \cos(2\mu_{3})\}  
\end{eqnarray}
One then optimizes over the remaining parameters in order to find $P_{s}$.  The result is shown in Fig.\ 2.  As expected, the success probability is lower than in the two-loop case, but, more interestingly, the behavior is quite different as well.  Instead of approaching a plateau, the success probability reaches a maximum and then decreases.  The success probability goes to $1/4$ at $\theta = \pi /4$, because at that value of $\theta$, the eight possible output states collapse down to two, $|\pm x\rangle$, so that each output state corresponds to four different trajectories.  The fact that the maximum success probability occurs at an intermediate value of $\theta$, where the final state depends on all of the measurement results, is more in line with one's expectations than the result in the two-loop case where $P_{s}$ was a maximum when it depended only on the result of the second measurement.

\section{Conclusion}
We have studied a number of instances of the effect of measurements on the final state of a quantum system, and our ability to use that state to retrodict the results of the measurements.  This ability can range from none to perfect, depending on the measurement and the initial state of the quantum system.  Using a qubit interferometer, we examined the retrodiction of a sequence of measurements for which we could vary the strength of the measurements.  The measurement we make on the final state of the quantum system depends on what we want to find out about the sequence of previous measurements.  We may want to find out the result of only one of the measurements, all of them, or find a measurement sequence that was not realized.

In our study of the two loop interferometer, we found that the highest success probability for determining the result of the first measurement occurred when both measurements were weaker than full projective measurements.  If the second measurement is a projective one, it erases the information about the first measurement.  Surprisingly, however, if we are trying to determine the results of both measurements, we found that the case with the highest success probability was when both measurements were projective.  We were also able to construct a measurement that would conclusively eliminate one of the trajectories.  In the case of a three loop interferometer, when determining the entire trajectory the highest success probability occurred when the measurements were weaker than projective measurements.

There are issues that could benefit from further study.  In all of the cases we examined, the party making the final state measurement and at least one of the parties making the earlier measurements share information, which means that this process can be viewed as a kind of communication channel.  This is the case, because the final state of a quantum system usually does carry information about the history of measurements on the system, and it is possible to gain access to this information by making measurements on the final state.  This suggests that the application of information measures to this problem would be a fruitful.  A second topic, which was not addressed here, is the role of the initial state.  Some initial states will prove better than others in transmitting the information about the measurement results to the final state.  We hope to make both of these issues the subject of future work.

\section*{Acknowledgment}
This research was supported by a grant from the John Templeton Foundation.

\section*{Appendix A}
Here we will look in more detail at the case of two two-outcome measurements when the second measurement is a projective one, which was discussed in Section III.  The strategy for determining the outcome of both measurements was first to measure which of the two subspaces, the one corresponding to $Q_{+}$ or the one corresponding to $Q_{-}$, the final state of the system is in.  Since it is definitely in one of these subspaces, and the subspaces are orthogonal, this measurement is deterministic.  One then performs one of two minimum-error measurements, which one depends on which subspace the state is in, in order to determine the result of the first measurement.  We want to determine the overall success probability of this procedure.

The probability that the final state is in the subspace corresponding to $Q_{+}$ is
\begin{equation}
P(Q_{+})=P(+,+) + P(+,-) ,
\end{equation}
and the probability that it is in the subspace corresponding to $Q_{-}$ is
\begin{equation}
P(Q_{-})=P(-,+) + P(-,-) ,
\end{equation}
where $P(j,k)=\| Q_{j}A_{k}\psi\|^{2}$, and $j,k\in \{ +,-\}$.  If we find that the state is in the subspace corresponding to $Q_{+}$ then we are faced with discriminating between two states, $|\psi^{++}_{out}\rangle = Q_{+}A_{+}|\psi\rangle /\|Q_{+}A_{+}\psi\|$, which occurs with a probability of
\begin{equation}
P(\psi_{out}^{++}|Q_{+})= \frac{P(+,+)}{P(+,+)+P(+,-)} ,
\end{equation}
and $|\psi_{out}^{+-}\rangle = Q_{+}A_{-}|\psi\rangle /\| Q_{+}A_{-}\psi\|$, which occurs with probability
\begin{equation}
P(\psi_{out}^{+-}|Q_{+})= \frac{P(+,-)}{P(+,+)+P(+,-)} .
\end{equation}
The success probability for this problem is given by $P_{s+}=(1/2)(1+\|\Lambda_{+}\|)$, where
\begin{eqnarray}
\Lambda_{+} & = &  \frac{1}{P(+,+)+P(+,-)}( Q_{+}A_{+}|\psi\rangle\langle\psi |A_{+}^{\dagger}Q_{+} 
\nonumber \\
& & -Q_{+}A_{-}|\psi\rangle\langle\psi |A^{\dagger}_{-}Q_{+})  .
\end{eqnarray}
Similarly, if one finds the final state in the support of $Q_{-}$, one wants to discriminate between $|\psi^{-+}_{out}\rangle = Q_{-}A_{+}|\psi\rangle /\|Q_{-}A_{+}\psi\|$ and $|\psi^{--}_{out}\rangle = Q_{-}A_{-}|\psi\rangle /\|Q_{-}A_{-}\psi\|$, and this can be done with a success probability of $P_{s-}=(1/2)(1+\|\Lambda_{-}\| )$, where
\begin{eqnarray}
\Lambda_{-} & = &  \frac{1}{P(-,+)+P(-,-)}( Q_{-}A_{+}|\psi\rangle\langle\psi |A_{+}^{\dagger}Q_{-} 
\nonumber \\
& & -Q_{-}A_{-}|\psi\rangle\langle\psi |A^{\dagger}_{-}Q_{-})  .
\end{eqnarray}
The overall success probability is
\begin{eqnarray}
P_{s} & = & (P(+,+)+P(+,-))P_{s+}\nonumber \\ 
& & +(P(-,+)+P(-,-))P_{s-} '
\end{eqnarray}
which is the same as Eq.\ (\ref{lambda1}).

Now we would like to show that the optimal four-element POVM for determining the final state in this case splits into a two-element POVM on the support of $Q_{+}$ and a two-element POVM on the support of $Q_{-}$.  This implies that the optimal POVM is the one discussed above, where we determine which subspace, support of $Q_{+}$ or support of $Q_{-}$, the final state is in and then apply the optimal two-element POVM to distinguish between the two possible final states in that subspace.  

Now suppose that the optimal POVM is $\{\Pi_{jk}|\, j,k=\pm\}$.  The success probability for this measurement is
\begin{equation}
P_{s}= \sum_{j,k=\pm} P(j,k)\langle\psi_{out}^{jk}|\Pi_{jk}|\psi_{out}^{jk}\rangle .
\end{equation}
We first note that
\begin{equation}
\sum_{j,k=\pm} Q_{+}\Pi_{jk}Q_{-} = \sum_{j,k=\pm} Q_{+}\Pi_{jk}Q_{-} = 0 ,
\end{equation}
which implies that
\begin{equation}
\label{id-sum}
\sum_{j,k=\pm} Q_{+}\Pi_{jk}Q_{+} + \sum_{j,k=\pm} Q_{-}\Pi_{jk}Q_{-} = I .
\end{equation}
Now define a new POVM
\begin{eqnarray}
\Pi_{++}^{\prime}=Q_{+}(\Pi_{++}+\Pi_{-+}+\Pi_{--})Q_{+} & \Pi_{+-}^{\prime}=Q_{+}\Pi_{+-}Q_{+} \nonumber \\
\Pi_{--}^{\prime} = Q_{-}(\Pi_{--}+ \Pi_{++}+\Pi_{+-}Q_{-} & \Pi_{-+}^{\prime}=Q_{-}\Pi_{-+}Q_{-}
\nonumber \\
\end{eqnarray}
This is a POVM, because its elements sum to the identity, see Eq.\ (\ref{id-sum}), and all of the operators are positive.  It also has the property that its success probability,
\begin{equation}
P_{s}^{\prime}= \sum_{j,k=\pm} P(j,k)\langle\psi_{out}^{jk}|\Pi_{jk}^{\prime}|\psi_{out}^{jk}\rangle .
\end{equation}
satisfies $P_{s}^{\prime}\geq P_{s}$.  For example, looking at the first terms in the sums for the two probabilities, we see that
\begin{eqnarray}
\langle\psi_{out}^{++}|\Pi_{++}^{\prime}|\psi_{out}^{++}\rangle & = & \langle\psi_{out}^{++}|\Pi_{++}|\psi_{out}^{++}\rangle \nonumber \\
& & + \langle\psi_{out}^{++}|(\Pi_{-+}+\Pi_{--})|\psi_{out}^{++}\rangle ,
\end{eqnarray}
where the second term on the right-hand side is clearly nonnegative.  So each term in the sum for $P_{s}^{\prime}$ is greater than or equal to the corresponding term in the sum for $P_{s}$.  Since we can take any POVM and create another one, which has the property that two of its elements have support in the support of $Q_{+}$ and two have support in the support of $Q_{-}$, and this second POVM has a greater than or equal success probability, the optimal POVM will have $\Pi_{++}$ and $\Pi_{+-}$ with support in the support of $Q_{+}$ and $\Pi_{-+}$ and $\Pi_{--}$ with support in the support of $Q_{-}$.

\section*{Appendix B}
The output states for the three-loop interferometer, in a form that is not normalized, are
\begin{eqnarray}
\label{3-loop-out}
|\tilde{\psi}^{+++}\rangle & = & \frac{1}{4}[ (\cos\theta - \sin (3\theta ))|+x\rangle + (\cos\theta + \sin\theta )|-x\rangle ] \nonumber \\
|\tilde{\psi}^{-++}\rangle & =& \frac{1}{4} [ (\cos (3\theta )-\sin\theta )|+x\rangle + (\cos\theta - \sin\theta )|-x\rangle ] \nonumber \\
|\tilde{\psi}^{++-}\rangle & = & \frac{1}{4} [ (\cos\theta - \sin\theta )|+x\rangle + (\cos (3\theta )-\sin\theta )|-x\rangle ] \nonumber \\ 
|\tilde{\psi}^{-+-}\rangle & = & \frac{1}{4} [ (\cos\theta + \sin\theta )|+x\rangle + (\cos\theta - \sin (3\theta )) |-x\rangle ] \nonumber \\
|\tilde{\psi}^{+-+}\rangle & = & \frac{1}{4} [ (\cos\theta - \sin\theta )|+x\rangle + (\cos\theta + \sin (3\theta )) |-x\rangle ]\nonumber \\
|\tilde{\psi}^{--+}\rangle & = & \frac{1}{4} [ (\cos\theta + \sin\theta )|+x\rangle + (\sin\theta + \cos (3\theta )|-x\rangle ] \nonumber \\
|\tilde{\psi}^{+--}\rangle & = & \frac{1}{4} [ (\sin\theta + \cos (3\theta ) )|+x\rangle + (\cos\theta + \sin\theta ) |-x\rangle ] \nonumber \\
|\tilde{\psi}^{---}\rangle & = & \frac{1}{4} [ (\cos\theta + \sin (3\theta )) |+x\rangle + (\cos\theta - \sin\theta ) |-x\rangle . \nonumber  \\
\end{eqnarray}
The vector $|\tilde{\psi}^{jkl}\rangle$, where $j,k,l\in \{ +,-\}$, is given by 
\begin{equation}
|\tilde{\psi}^{jkl}\rangle = HA_{j}HA_{k}HA_{l}|+x\rangle .
\end{equation}
The probability that the output state is $|\tilde{\psi}^{jkl}\rangle /\|\tilde{\psi}^{jkl}\|$ is just $\|\tilde{\psi}^{jkl}\|^{2}$.

\end{document}